\documentclass[12pt]{article} 

\usepackage{graphicx}
\usepackage{amsmath,amssymb,amsbsy}
\usepackage{color,colortbl}
\usepackage{cite}

\usepackage{multirow}
\usepackage{fullpage}
\usepackage{tabularx}
\usepackage{booktabs} 

\newcommand\bm\boldsymbol 

\begin{document}

\title{
Ratios of Hidden-Charm Compact Pentaquark Decay \\ Widths in Quark-Diquark Model
}

\author{A.\,A.~Dobrynina, A.\,Ya.~Parkhomenko, A.\,V.~Zinchenko \\ 
P.\,G.~Demidov Yaroslavl State University, Yaroslavl, Russia 
}


\date{\today}

\maketitle

\begin{abstract}
A number of resonances comparable with a hypothesis of hidden-charm pentaquark is observed 
by the LHCb Collaboration. We interpret these narrow resonances as compact hidden-charm 
diquark-diquark-antiquark systems. Within this assumption, an interplay between the 
charmonium and open-charm modes is considered. Ratios of such modes for non-strange 
pentaquarks are obtained and discussed. 
\end{abstract}

\section{Introduction}

\begin{figure}
\begin{center} 
\includegraphics[width=0.55\textwidth]{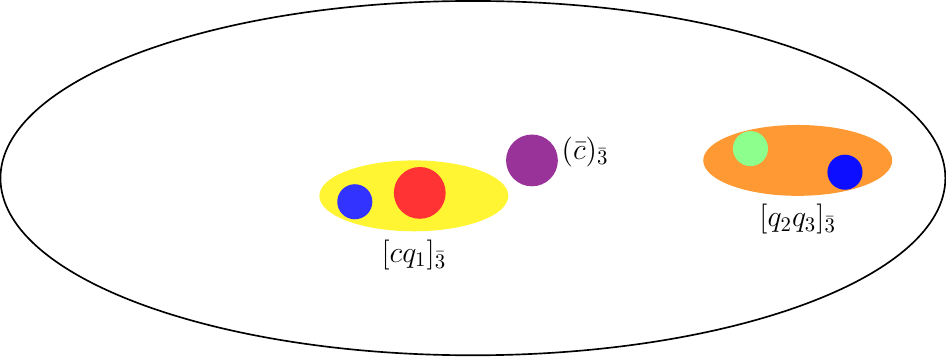}  
\end{center}
\caption{
The hidden-charm pentaquark in the diquark-diquark-antiquark model 
used for getting the mass spectrum in~\cite{Ali:2019npk, Ali:2019clg}. 
}
\label{fig:pentaquark-DDQ-model}
\end{figure}

At present, production, properties, and decays of bottom baryons 
are intensively studied both experimentally and theoretically.  
Of special interest are $\Lambda_b$-, $\Xi_b^0$- and $\Xi_b^-$-baryons  
which are decaying weakly and many decay modes are found experimentally~\cite{ParticleDataGroup:2022pth}.  
$\Lambda_b$-baryon is a bound state of heavy $b$-quark and a pair of light $u$- and $d$-quarks. 
Its mass and lifetime are $m_{\Lambda_b} = 5619.51 \pm 0.23~\text{\rm MeV}$ and 
$\tau_{\Lambda_b} = \left ( 1.466 \pm 0.010 \right ) \times 10^{-12}~\text{\rm sec}$, 
respectively,~\cite{ParticleDataGroup:2022pth}, and such a large lifetime is due to 
weak interactions.    
More than 40 decay modes with branching fractions exceeded $10^{-6}$ are experimentally 
found~\cite{ParticleDataGroup:2022pth}. Two exotic resonances, $P_\psi^N (4380)^+$ and 
$P_\psi^N (4450)^+$, consistent with the pentaquark interpretation were originally found 
in the $\Lambda_b \to p + J/\psi  + K^-$ decay by the LHCb Collaboration~\cite{Aaij:2015tga}.  
Later in the same channel on higher statistics, the LHCb found three narrow resonances: 
$P_\psi^N (4312)^+$, $P_\psi^N (4440)^+$, and $P_\psi^N (4457)^+$, while the existence 
of the broad one, $P_\psi^N (4380)^+$, remains under question~\cite{LHCb:2019kea}. 
The evidence of the original pentaquark resonances was also announced in the 
$\Lambda_b \to p + \pi^- + J/\psi$ decay by the LHCb Collaboration~\cite{Aaij:2016ymb}. 
The evidence of the resonance consistent with the strange $P_{\psi s}^\Lambda (4459)^0$ 
pentaquark was reported by the LHCb Collaboration~\cite{LHCb:2020jpq} in the 
$\Xi_b^- \to \Lambda + J/\psi + K^-$ decay of the $\Xi_b^-$-baryon, the $SU(3)_F$-partner 
of $\Lambda_b$. Unfortunately, spin-parities of all these resonances are not yet determined 
and theoretical speculations about their quantum numbers and binding mechanisms are still 
debatable (see, for example, the latest reviews on this topic~\cite{Ali:2019roi, Brambilla:2019esw, Mai:2022eur}).  
Note that several dynamical models of pentaquarks are suggested: baryon-meson model 
(molecular pentaquark), triquark-diquark model, diquark-diquark-antiquark model, etc. 
For example, in the diquark-diquark-antiquark model~\cite{Ali:2019npk, Ali:2019clg}, 
dynamics is determined by interaction of light diquark $[q_2 q_3]$, heavy diquark $[c q_1]$ 
and $c$-antiquark, where $q_i$ is one of the light~$u$-, $d$- or $s$-quarks as shown 
in Fig.~\ref{fig:pentaquark-DDQ-model}. As far as the calculation of the mass spectrum 
in this model done and experimentally observed resonances can be successfully identified  
with theoretically calculated states, hidden-charm pentaquark decay mechanism is not 
working out completely. Here, we give arguments and qualitative estimates of a possible 
mechanism similar to one suggested for decays of hidden-charm tetraquarks in~\cite{Maiani:2017kyi}. 

\section{Double Well Potential in Tetraquarks}

L.~Maiani, A.\,D.~Polosa, and V.~Riquer in~\cite{Maiani:2017kyi} suggested the hypothesis: 
a tetraquark can plausibly be represented by two diquarks in double well potential 
separated by a barrier. In this case, 
there are two length scales: the diquark radius~$R_{Qq}$ and tetraquark radius~$R_{4q}$ 
which are assumed to be well separated and their ratio can be estimated as 
$\lambda = R_{4q}/R_{Qq} \geq 3$.  
Tunneling transitions of quarks result into tetraquark strong decays. They have also claimed 
that the diquark radius $R_{Qq}$ in tetraquark can be different from the diquark radius 
$R_{Qq}^{\rm baryon}$ in baryon. An increase of the experimental resolution and statistics 
is crucial to support or disprove this hypothesis.  

Let us start from the decays of hidden-charm tetraquarks to two $D$-mesons based 
on the $X (3872)$ as an example. Diquark-antidiquark system, $([c q] [\bar c \bar q])$ 
can rearrange itself into a pair of color singlets by exchanging quarks through 
tunneling transition. Small overlap between constituent quarks in different wells 
suppresses the quark-antiquark pair from the direct annihilation. 
So, the two stage process should occur within this mechanism: first, the light 
quark and antiquark switch of among two wells and, second, the quark-antiquark 
pairs obtained are evolved in their color-singlet components (two $D$-mesons).    
Including diquark spins (subscripts), consider the states~\cite{Maiani:2017kyi}:  
\begin{equation}   
\Psi^{(1)}_{\cal D} = [cu]_0 (x)\, [{\bar c\bar u}]_1 (y), 
\quad 
\Psi^{(2)}_{\cal D} = {\cal C} \Psi^{(1)}_{\cal D} = [cu]_1 (y)\, [{\bar c \bar u}]_0 (x) , 
\label{eq:Psi-D}
\end{equation}   
with ${\cal C}$ being the charge conjugation operator.   
After Fierz rearrangements of color and spin indices and assuming quarks 
to be non-relativistic particles, in evident meson notations one obtains: 
\begin{eqnarray}
&& 
\Psi^{(1)}_{\cal D} = A\, D^0 {\bar{\bm D}}^{*0} 
- B\, {\bm D}^{*0} {\bar D}^0 
+ i C\, {\bm D}^{*0}{\bm \times}{\bar{\bm D}}^{*0} , 
\nonumber \\ 
&& 
\Psi^{(2)}_{\cal D} = B\, D^0 {\bar{\bm D}}^{*0} 
- A\, {\bm D}^{*0} {\bar D}^0 
- i C\, {\bm D}^{*0}{\bm \times}{\bar{\bm D}}^{*0} , 
\nonumber 
\end{eqnarray}
where $A$, $B$, and $C$ are non-perturbative coefficients associated to barrier 
penetration amplitudes for different total spins of $u$ and $\bar u$ light quarks.   

The other possible decay channel of hidden-charm tetraquarks is to a charmonium 
and light meson. The tunneling transition of light quarks is as follows: 
\begin{equation}
X_u \sim \frac{1}{\sqrt 2} \left [ \Psi^{(1)}_{\cal D} + \Psi^{(2)}_{\cal D} \right ] 
= \frac{A+B}{\sqrt 2} \left [ D^0 {\bar{\bm D}}^{*0} - {\bm D}^{*0} {\bar D}^0 \right ] , 
\label{xuprime}
\end{equation} 
while the tunneling transition of heavy quarks with finite masses: 
\begin{equation}
X_u \sim a\,i {\bm J/\bm\psi}{\bm \times} \left ( {\bm \omega} + {\bm \rho}^0 \right ) . 
\label{xus}
\end{equation}
For the tunneling amplitude in the leading semiclassical approximation, one has   
${\cal A}_M \sim e^{-\sqrt{2ME}\ell}$, where $E$ and $\ell$ are the barrier height and extension. 
For the constituent quark masses, $m_q$ and $m_c$, $E = 100$~MeV and 
$\ell = 2$~fm~\cite{Maiani:2017kyi}, one can estimate the ratio of amplitudes squared to be:    
\begin{equation}
R = \left [ a/(A + B) \right ]^2 
\sim \left ( {\cal A}_{m_c}/{\cal A}_{m_q} \right )^2 \sim 10^{-3} . 
\label{eq:R-tetraquarks}
\end{equation}
With the decay momenta $p_\rho \simeq 124$~MeV and $p_D \simeq 2$~MeV~\cite{Maiani:2017kyi}, 
the decay width ratio has the following estimate:   
\begin{equation}
\frac{\Gamma (X(3872)\to J/\psi\, \rho)}{\Gamma (X(3872) \to D\, \bar D^*)} 
= \frac{p_\rho}{p_D}\, R \sim 0.1 . 
\end{equation}
Its comparison with existing experimental data~\cite{ParticleDataGroup:2022pth}: 
\begin{equation}
B_{\rm exp} (X(3872) \to J/\psi\, \rho) = (3.8 \pm 1.2)\% , 
\qquad      
B_{\rm exp} (X(3872) \to D\, \bar D^*) = (37 \pm 9)\% , 
\label{eq:B-exp-X3872}
\end{equation}
shows the excellent agreement, $R_{\rm exp} \simeq 0.1$, but one should remember 
that the coefficients associated to barrier penetration amplitudes are 
non-perturbative quantities and require a more detail information about 
a potential shape and parameters entering the potential. 

\begin{figure}
\begin{center} 
\includegraphics[width=0.45\textwidth]{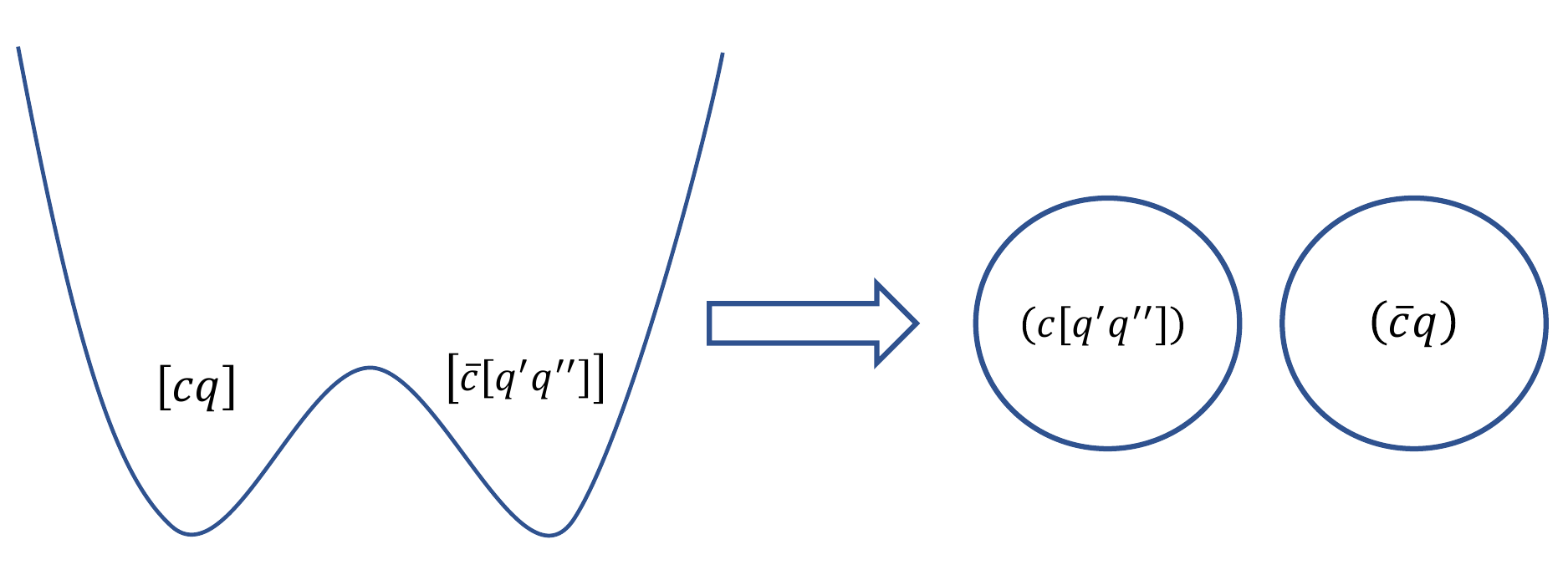}  
\end{center} 
\caption{
The hidden-charm pentaquark decay to the charmed baryon and charmed meson.  
}
\label{fig:pentaquark-decay-to-OCH}
\end{figure}

\begin{figure}
\begin{center} 
\includegraphics[width=0.45\textwidth]{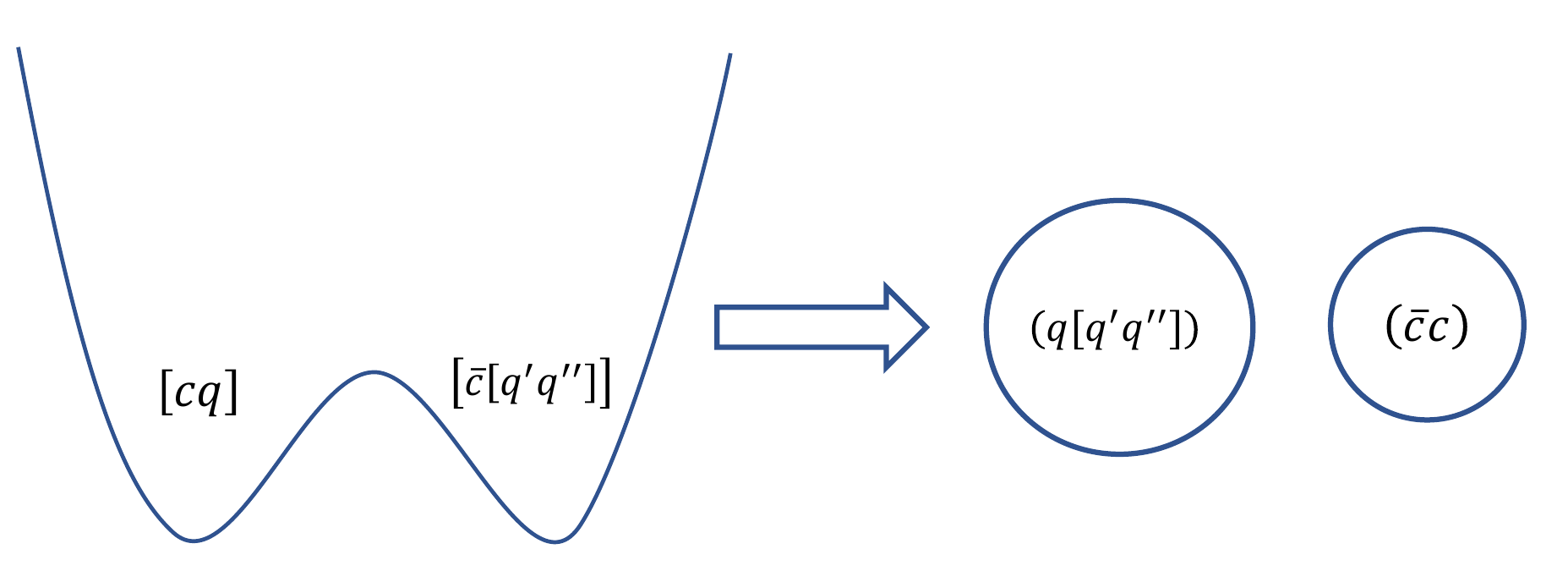}  
\end{center} 
\caption{
The hidden-charm pentaquark decay to the charmonium and light baryon.  
}
\label{fig:pentaquark-decay-to-charmonium}
\end{figure}

\section{Double Well Potential in Pentaquarks} 

In case of pentaquarks, similar hypothesis can be formulated: a pentaquark can 
be represented by the heavy diquark and heavy triquark in double well potential 
separated by barrier~\cite{Ali:2019clg} as shown in Figs.~\ref{fig:pentaquark-decay-to-OCH} 
and~\ref{fig:pentaquark-decay-to-charmonium}. 
There are two triquark-diquark representations:  
\begin{eqnarray} 
\Psi^D_1 = \frac{1}{\sqrt 3} \left [ \frac{1}{\sqrt 2} \, \epsilon_{ijk} \bar c^i 
\left [ \frac{1}{\sqrt 2} \, \epsilon^{jlm} c_l q_m \right ] \right ] 
\left [ \frac{1}{\sqrt 2} \, \epsilon^{knp} q^\prime_n q^{\prime\prime}_p \right ] 
\equiv \left [ \bar c \left [ c q \right ] \right ]  \left [ q^\prime q^{\prime\prime} \right ] ,  
\label{eq:PsiD1-def} \\ 
\Psi^D_2 = \frac{1}{\sqrt 3} \left [ \frac{1}{\sqrt 2} \, \epsilon_{ikj} \bar c^i 
\left [ \frac{1}{\sqrt 2} \, \epsilon^{knp} q^\prime_n q^{\prime\prime}_p \right ] \right ]  
\left [ \frac{1}{\sqrt 2} \, \epsilon^{jlm} c_l q_m \right ] 
\equiv \left [ \bar c \left [ q^\prime q^{\prime\prime} \right ] \right ] \left [ c q \right ] ,  
\label{eq:PsiD2-def}
\end{eqnarray} 
where all the diquarks are assumed to be $\bar 3$-color states. 
From the color algebra, these states are related, $\Psi^D_2 = - \Psi^D_1$,  
but other internal dynamical properties can be different.  
The color connection of quarks in $\Psi^D_1$ is used for getting the mass spectrum in~\cite{Ali:2019clg}.   
The color structure of $\Psi^D_2$ is suitable for study the pentaquark strong decays. 
This is employed in the Dynamical Diquark Model  
of multiquark exotic hadrons~\cite{Lebed:2015tna,Lebed:2017min,Giron:2019bcs}. 
The color-singlet combinations are meson-baryon alternatives:  
\begin{eqnarray*} 
&& 
\Psi^H_1 = 
\left ( \frac{1}{\sqrt 3} \, \bar c^i c_i \right )
\left [ \frac{1}{\sqrt 6} \, \epsilon^{jkl} q_j q^\prime_k q^{\prime\prime}_l \right ] 
\equiv \left ( \bar c c \right ) \left [ q q^\prime q^{\prime\prime} \right ] ,  
\label{eq:PsiH1-def} \\ 
&& 
\Psi^H_2 = 
\left ( \frac{1}{\sqrt 3} \, \bar c^i q_i \right )
\left [ \frac{1}{\sqrt 6} \, \epsilon^{jkl} c_j q^\prime_k q^{\prime\prime}_l \right ]  
\equiv \left ( \bar c q \right ) \left [ c q^\prime q^{\prime\prime} \right ] ,  
\label{eq:PsiH2-def} \\ 
&& 
\Psi^H_3 = 
\left ( \frac{1}{\sqrt 3} \, \bar c^i q^\prime_i \right )
\left [ \frac{1}{\sqrt 6} \, \epsilon^{jkl} c_j q_k q^{\prime\prime}_l \right ]   
\equiv \left ( \bar c q^\prime \right ) \left [ c q q^{\prime\prime} \right ] , 
\label{eq:PsiH3-def} \\  
&& 
\Psi^H_4 = 
\left ( \frac{1}{\sqrt 3} \, \bar c^i q^{\prime\prime}_i \right )
\left [ \frac{1}{\sqrt 6} \, \epsilon^{jkl} c_j q_k q^\prime_l \right ]  
\equiv \left ( \bar c q^{\prime\prime} \right ) \left [ c q q^\prime \right ] .   
\label{eq:PsiH4-def}  
\end{eqnarray*} 
From these four states, two of them, $\Psi^H_1$ and $\Psi^H_2$, only 
satisfy the heavy-quark-symmetry condition~\cite{Ali:2019clg}.  
The light $\left [ q^\prime q^{\prime\prime} \right ]$-diquark is transmitted intact, 
retaining its spin quantum number, from the $b$-baryon to pentaquark.
Keeping the color of the light diquark unchanged, a convolution  
of two Levi-Civita tensors entering the triquark gives: 
\begin{equation} 
\Psi^D_1 = - \frac{\sqrt 3}{2} \left [ \Psi^H_1 + \Psi^H_2 \right ] .   
\label{eq:PsiD1-PsiH1-PsiH2}
\end{equation}
The color reconnection is not enough to reexpress the pentaquark operator  
as a direct product of the meson and baryon operators. 
Spins of quarks and diquarks should be projected onto the definite hadronic spin states. 
One needs to know the Dirac structure of pentaquark operators  
to undertake the Fierz transformations in the Dirac space under 
assumption that quarks are non-relativistic. 
Let us exemplify this by considering the $P_\psi^N (4312)^+$ pentaquark.   
Diquark-diquark-antiquark operators with spinless heavy and light diquarks are~\cite{Ali:2019clg}:  
%
\begin{eqnarray} 
&& 
\Psi_1^{H(1)} (x, y) = \frac{1}{3}
\left ( \tilde c^i (x)\, \sigma_2 \right ) 
\left ( c_i (y)\, \sigma_2\, q_k (y) \right ) d_0^k (x) ,  
\label{eq:psiH1-1-spin} \\ 
&& 
\Psi_2^{H(1)} (x, y) = \frac{1}{3} 
\left ( \tilde c^i (x)\, \sigma_2 \right ) 
\left ( c_k (y)\, \sigma_2\, q_i (y) \right ) d_0^k (x) .  
\label{eq:psiH2-1-spin}  
\end{eqnarray}
For the lowest lying pentaquark, $q = u$ and $d_0 = \left [ u\, C\, \gamma_5\, d \right ]$, 
being scalar diquark. 
For simplicity, all the quarks are considered in the non-relativistic limit.  
%
%
%
%
%
%
After the Fierz transformation of the Pauli matrices and suppressing 
position dependence of the fields, they can be rewritten in terms of hadrons:  
\begin{equation} 
\Psi_1^{H(1)} = - \frac{i}{\sqrt 2} \left [ 
a\, \eta_c + b \left ( \boldsymbol\sigma\, \boldsymbol{J}\boldsymbol{/\psi} \right ) 
\right ] p , 
\quad 
\Psi_2^{H(1)} = - \frac{i}{\sqrt 2} \left [ 
A\, \bar D^0 + B \left ( \boldsymbol\sigma\, \bar{\boldsymbol{D}}^{*0} \right ) 
\right ] \Lambda_c^+ .   
\end{equation}
Here, $A$ and~$B$ ($a$ and~$b$) are non-perturbative coefficients associated 
with barrier penetration amplitudes for the light (heavy) quark. 
They are equal in the limit of the naive Fierz coupling.  
The decays of the pentaquark into the $D$-meson and charmed baryon and into 
a charmonium and light baryon through the tunneling transition are shown 
in Figs.~\ref{fig:pentaquark-decay-to-OCH} and~\ref{fig:pentaquark-decay-to-charmonium}.

Similarly, diquark-diquark-antiquark operators containing heavy diquark 
with the spin $S_{hd} = 1$ and light diquark with $S_{ld} = 0$: 
\begin{eqnarray} 
&& 
\boldsymbol\Psi_1^{H(2)} (x, y) = \frac{1}{3}
\left ( \tilde c^i (x)\, \sigma_2 \right ) 
\left ( c_i (y)\, \sigma_2\, \boldsymbol\sigma\, q_k (y) \right ) d_0^k (x) ,  
\label{eq:psiH1-2-spin} \\ 
&& 
\boldsymbol\Psi_2^{H(2)} (x, y) = \frac{1}{3} 
\left ( \tilde c^i (x)\, \sigma_2 \right ) 
\left ( c_k (y)\, \sigma_2\, \boldsymbol\sigma\, q_i (y) \right ) d_0^k (x) .  
\label{eq:psiH2-2-spin}  
\end{eqnarray}
Being direct product of spinor and vector, they need to be separated 
into two states with spins $J = 1/2$ and $J = 3/2$.   
For $P_\psi^N (4312)^+$ interpreted as $J^P = 3/2^-$ pentaquark~\cite{Ali:2019npk, Ali:2019clg}, 
decompositions in term of hadrons are as follows:   
\begin{eqnarray} 
&& 
\boldsymbol\Psi_1^{H(3/2)} = \frac{i \sqrt 2}{3} \left \{ 
b'\, \boldsymbol{J/\psi} - 2 i c' \left [ \boldsymbol\sigma \times \boldsymbol{J/\psi} \right ]  
\right \} p ,  
\label{eq:psiH1-3/2-spin-h-uf} \\ 
&& 
\boldsymbol\Psi_2^{H(3/2)} = - \frac{i \sqrt 2}{3} \left \{ 
B'\, \bar{\boldsymbol{D}}^{*0} - 2 i C' \left [ \boldsymbol\sigma \times \bar{\boldsymbol{D}}^{*0} \right ]  
\right \} \Lambda_c^+ .   
\label{eq:psiH2-3/2-spin-h-uf}  
\end{eqnarray}
%
So, $P_\psi^N (4312)^+$ is mainly decaying either to $J/\psi\, p$ final state,  
in which it was observed, or to $\Lambda_c^+\, \bar D^{*0}$.  

The tunneling amplitude in leading semiclassical approximation, has a similar  
exponential behavior as for tetraquarks: ${\cal A}_M \sim e^{-\sqrt{2ME} \ell}$, 
where $E$ and $\ell$ are barrier height and extension.  
For constituent quark masses, $m_u$ and $m_c$, and keeping the same values 
as for tetraquarks, $E = 100$~MeV and $\ell = 2$~fm~\cite{Maiani:2017kyi}, 
the ratio of amplitudes squared has the same order of magnitude as~(\ref{eq:R-tetraquarks}):    
\begin{equation}
R_{\rm penta} = \frac{|b'|^2 + 4 |c'|^2}{|B'|^2 + 4 |C'|^2} 
\sim \left ( \frac{{\cal A}_{m_c}}{{\cal A}_{m_u}} \right )^2 \sim 10^{-3} \sim R . 
\label{eq:R-pentaquarks}
 \end{equation}
With the decay momenta $p_p \simeq 660$~MeV and $p_{\Lambda_c} \simeq 200$~MeV, 
being comparable to each other, one can get the ratio of pentaquark decay widths:   
\begin{equation}
\frac{\Gamma (P_\psi^N (4312)^+ \to J/\psi\, p)}
     {\Gamma (P_\psi^N (4312)^+ \to \Lambda_c^+\, \bar D^{*0})} 
= \frac{p_p}{p_{\Lambda_c}}\, R_{\rm penta} \sim 10^{-3} . 
\end{equation}
If this approach is correct, $P_\psi^N (4312)^+$ should be also searched  
in $\Lambda_b^0 \to \Lambda_c^+\, \bar D^{*0}\, K^-$ decay with good chances 
to be observed. This can also be applied to decays of the 
$P_{\psi s}^\Lambda (4459)^0$ pentaquark which we left for a future publication.

\section{Conclusions} 

The Quark-Diquark approach used for pentaquarks is working quite successful 
in predictions of masses of heavy baryons and doubly-heavy exotic hadrons.  
Decay width of tetraquarks with hidden charm or bottom can be explained 
within the quark-diquark model by a presence of a barrier between heavy 
diquark and antidiquark. 
Similarly, decay width of pentaquarks with hidden charm or bottom can 
be explained within the quark-diquark model by a presence of a barrier 
between heavy diquark and heavy triquark.  
If this approach is correct, $P_\psi^N (4312)^+$-pentaquark should be also 
searched in the $\Lambda_b^0 \to \Lambda_c^+\, \bar D^{*0}\, K^-$ decay mode 
with good chances to be found.

\section*{Acknowledgments}
AP would like to thank Prof. Ahmed Ali for useful discussions. 
A.\,D. and A.\,P. are supported by the Russian Science Foundation  
(Project No.~22-22-00877, https://rscf.ru/project/22-22-00877/). 
A,\,Z. is supported by the Russian Foundation for Basic Research  
(Project \textnumero~20-32-90205).

%
%

%


\begin{thebibliography}{99}
%
\bibitem{ParticleDataGroup:2022pth}
Particle Data Group (R.~L.~Workman \emph{et al.}), 
PTEP \textbf{2022}, 083C01 (2022). 
%
\bibitem{Aaij:2015tga}          
LHCb Collab. (R.~Aaij \emph{et al.}), 
Phys. Rev. Lett. \textbf{115}, 072001 (2015)    
[arXiv:1507.03414 [hep-ex]]. 
%
\bibitem{LHCb:2019kea}
LHCb Collab. (R.~Aaij \emph{et al.}), 
Phys. Rev. Lett. \textbf{122}, 222001 (2019)  
[arXiv:1904.03947 [hep-ex]]. 
%
\bibitem{Aaij:2016ymb}
LHCb Collab. (R.~Aaij \emph{et al.}), 
Phys. Rev. Lett. \textbf{117}, 082003 (2016)   
[Addendum: Phys. Rev. Lett. \textbf{118}, 119901 (2017)]  
[arXiv:1606.06999 [hep-ex]]. 
%
\bibitem{LHCb:2020jpq}
LHCb Collab. (R.~Aaij \emph{et al.}), 
Sci. Bull. \textbf{66}, 1278 (2021)  
[arXiv:2012.10380 [hep-ex]]. 
%
\bibitem{Ali:2019roi}
A.~Ali, L.~Maiani, and A.~D.~Polosa, 
\emph{Multiquark Hadrons} (Cambridge University Press, 2019). 
%
\bibitem{Brambilla:2019esw}
N.~Brambilla, S.~Eidelman, C.~Hanhart, A.~Nefediev, C.~P.~Shen, C.~E.~Thomas, A.~Vairo, and C.~Z.~Yuan,
Phys. Rept. \textbf{873}, 1 (2020)
[arXiv:1907.07583 [hep-ex]]. 
%
\bibitem{Mai:2022eur}
M.~Mai, U.~G.~Mei\ss{}ner and C.~Urbach,
Phys. Rept. \textbf{1001}, 1 (2023)
[arXiv:2206.01477 [hep-ph]]. 
%
\bibitem{Ali:2019npk}            
A.~Ali and A.~Y.~Parkhomenko,
Phys. Lett. B \textbf{793}, 365-371 (2019)
[arXiv:1904.00446 [hep-ph]].
%
\bibitem{Ali:2019clg}
A.~Ali, I.~Ahmed, M.~J.~Aslam, A.~Y.~Parkhomenko, and A.~Rehman,
JHEP \textbf{10}, 256 (2019)
[arXiv:1907.06507 [hep-ph]].
%
\bibitem{Maiani:2017kyi}
L.~Maiani, A.~D.~Polosa, and V.~Riquer,
Phys. Lett. B \textbf{778}, 247 (2018)
[arXiv:1712.05296 [hep-ph]].
%
\bibitem{Lebed:2015tna}
R.~F.~Lebed,
Phys. Lett. B \textbf{749}, 454 (2015)
[arXiv:1507.05867 [hep-ph]]. 
%
\bibitem{Lebed:2017min}
R.~F.~Lebed,
Phys. Rev. D \textbf{96}, 116003 (2017)
[arXiv:1709.06097 [hep-ph]]. 
%
\bibitem{Giron:2019bcs}
J.~F.~Giron, R.~F.~Lebed, and C.~T.~Peterson,
JHEP \textbf{05}, 061 (2019)
[arXiv:1903.04551 [hep-ph]].
%
\end{thebibliography}
\end{document}